\documentclass{amsart}
\usepackage{cite}
\def\R{{\mathbb R}}
\def\C{{\mathbb C}}
\def\N{{\mathbb N}}
\def\Z{{\mathbb Z}}
\def\eps{\varepsilon}
\def\alphae{\alpha^{\eps}}
\def\ue{u^{\eps}}
\def\hatalphae{{\hat\alpha}^{\eps}}
\def\xe{x^{\eps}}
\def\lame{\lambda^{\eps}}
\def\Max{\mathop{\textup{Max}}\limits}
\def\M{\mathfrak{M}}
\def\II{\mathfrak{I}}
\def\wei{{w_i^{\eps}}}
\newtheorem{theorem}{Theorem}[section]
\newtheorem{corollary}[theorem]{Corollary}

\newtheorem{definition}[theorem]{Definition}

 \makeatletter
\def\@oddhead{\underline{\hbox to \textwidth{Negativity Not Problem For Ultra-discrete Limit\hfill A. Kasman/S. Lafortune}}}
\let\@evenhead\@oddhead \def\@oddfoot{$\overline{\hbox to
\textwidth{\hfill \bf\thepage\hfill}}$} \let\@evenfoot\@oddfoot
\advance\footskip.5in \makeatother

\title{When is negativity not a problem for the ultra-discrete limit?}

\author{Alex Kasman}

\address{Department of Mathematics / College of Charleston}

\email{kasmana@cofc.edu}

\author{St\'ephane Lafortune}

\address{Department of Mathematics / College of Charleston}

\email{lafortunes@cofc.edu}
\begin{document}

\begin{abstract}
The `ultra-discrete limit' has provided a link between integrable difference equations and cellular automata displaying soliton like solutions.  In particular, this procedure generally turns \textit{strictly positive} solutions of algebraic difference equations with \textit{positive} coefficients into corresponding solutions to equations involving the ``Max" operator.  
Although it certainly is the case that dropping these positivity conditions creates potential difficulties, it is still possible for solutions to persist under the ultra-discrete limit even in their absence.  To recognize when this will occur, one must consider whether a certain expression, involving a measure of the rates of convergence of different terms in the difference equation and their coefficients, is equal to zero. 
Applications discussed include the solution of elementary ordinary difference equations, a discretization of the Hirota Bilinear Difference Equation and the identification of integrals of motion for ultra-discrete equations.
\end{abstract}

\maketitle
\section{Introduction}

Integrable nonlinear partial differential equations are of special interest in part because of their  `soliton solutions' which exhibit particle-like behavior \cite{BKY,GGKM,KdV95,ZK}.  The link between these equations and the cellular automata that exhibit superficially similar solutions \cite{FPS,NTSWK,PST,TS} was the observation in \cite{TTMS} that solutions to the discretizations of the soliton equations become solutions to these cellular automata under the `ultra-discrete limit'.
Loosely speaking, this works because of the fact that for $a,b\in\R$
\begin{equation}
\lim_{\eps\to0^+} \eps\ln\left(e^{a/\eps}+e^{b/\eps}\right)=\Max(a,b).
\label{orig-max}
\end{equation}
This formula can be understood intuitively by noting that the term corresponding to the smaller of the two exponents will be relatively insignificant when $\eps$ is very small and that with it eliminated the entire expression is simply equal to the numerator of the other exponent.  It can also be easily verified using L'H\^opital's rule.  

In applying \eqref{orig-max} to difference equations, previous authors have been very cautious allowing negative coefficients in front of the exponentials, and have not been concerned at all about the question of how allowing $a$ and $b$ in the formula above to depend on $\eps$ might affect the outcome. 
In fact, they are correct that letting those coefficients be negative may invalidate the prediction that the resulting limit will merely be given by the ``$\Max$".  In this paper, we will be especially interested in the case in which the linear combination of exponential functions \textit{does} involve negative coefficients.  Our viewpoint is that it is worthwhile to investigate this situation to understand what can go wrong.  

Theorem~\ref{main} demonstrates two surprising facts:  (a) Even if negative coefficients are involved, a formula analogous to \eqref{orig-max} continues to work ``most of the time''. (b) In determining when it might fail it is necessary to explicitly consider the rate of convergence of the exponents under the limit in $\eps$.  This theorem is illustrated by an application to two simple, ordinary difference equations in Section~\ref{sec:ordinary}.  In a more sophisticated application, the theorem is used to guarantee the production of a solution to an ultra-discrete limit of the Hirota Bilinear Difference Equation in Section~\ref{sec:hbde}.  Finally, these results are applied to the question of when integrals of motion for difference equations survive the ultra-discrete limit in Section~\ref{sec:integrals}.

Both to refer readers to other work that may be of interest and to emphasize the differences with this paper, we would like to conclude this introduction with a discussion of three other papers.
Equation \eqref{orig-max} has been previously generalized in \cite{YaNaAs05} to the case in which $a$ and $b$ are complex constants.  This is of great interest, but clearly quite different from what we do below by allowing the argument of the logarithm to be an arbitrary linear combination of exponentials and letting the exponents be real functions of $\eps$.  Also, \cite{IGRS} presents work related to the goals of the present paper by addressing the question of how one can take an ultra-discrete limit of a solution to a difference equation which takes on negative values.   Their approach involves an alternative method of introducing dependence on $\eps$ into the solution so that the new variable can be real even in the case that the original variable was negative.  However, in doing so they are careful to be certain that they always apply \eqref{orig-max} only in the case that the coefficients of the exponentials are equal to one and make no explicit mention of the possibility that the results may be affected by the dependence of the exponents on $\eps$.  So, that paper takes what might be viewed as an ``orthogonal'' approach to the question to the one that will be pursued below.  It might be of interest to attempt to combine the two, by applying Theorem~\ref{main} to the {\sl{s-ultra-discrete limit}} as defined in that paper.  Finally, perhaps most closely related to this paper in spirit, \cite{ON} generalizes the procedure of ultra-discretization by formally extending the Max-Plus algebra through the introduction of an extra $\Z_2$ component.  This extension of the Max-Plus semi-ring to a full ring, as is normally done in the Max-Plus community by moving to the symmetrized ring $\mathbb{S}_{max}$ \cite{Gaubert}, means that the resulting equations are not ordinary ultra-discrete equations (the $\Max$ operator and even equality have slightly different meanings).  In contrast, in this paper we address the question of when negativity fails to create any problems under the original definition of the ultra-discrete limit and integrable ultra-discrete equations.

\subsection{Notation}

In the following, 
%
a superscript of $\eps$ will indicate that the parameter depends \textit{differentiably} on the real variable $\eps$ in some open interval interval $(0,\beta)$.  Thus $u_n^{\eps}$ would be an $\eps$ dependent bi-infinite sequence and could be a solution to a difference equation in which the coefficients also depend on $\eps$, and we would only call it a solution if the equation were true not only for all $n$ but also for all $\eps\in (0,\beta)$.  (E.g. $u_n^{\eps}=\eps n$ satisfies $u_{n+1}^{\eps}u_{n-1}^{\eps}=(u_n^{\eps})^2-\eps^2$).

\subsection{An Enlightening Example}\label{sec:examp}

As we will see in Theorem~\ref{main} below, it is possible to generalize \eqref{orig-max} to the case in which the argument of the logarithm is a more general linear combination of exponentials whose exponents depend on $\eps$.  As one might guess, the value of the limit turns out to be equal to the maximum of the limits of the numerators of the exponents.  However, the theorem can only be applied subject to the condition that a certain expression involving the coefficients and the rates of convergence is not equal to zero.  
Following is a cautionary example whose purpose is to illustrate the fact that failing to consider this condition is liable to produce incorrect results.

If we consider the expression
$$
\tau=e^{\ue_1/\eps}+\gamma e^{\ue_2/\eps}
$$ 
then using the previously stated argument that the term with the lower power can simply be ignored, we may naively think that 
\begin{equation}
\lim_{\eps\to0^+}\eps\ln\tau=\Max(u_1,u_2)
\label{temp}
\qquad
\hbox{
where } \qquad
u_i=\lim_{\eps\to0^+}\ue_i.  
\end{equation}
However, \eqref{temp} may be false for some choices of $\ue_i$ and $\gamma$.

For instance, if we let 
$$
\ue_1=a+2\eps\ln\left(e^{-k/\eps}+\sqrt{8+e^{-2k/\eps}}\right)
\qquad
\ue_2=a+\eps\ln2\qquad
\hbox{and}
\qquad
\gamma=-4
$$
where $k>0$ and $a$ is arbitrary, then
$$
u_1=u_2=a.
$$
However, despite the predictions of equation \eqref{temp}, one can check that in fact 
$$
\lim_{\eps\to0^+}\eps\ln\tau=a-k.
$$

%

The total avoidance of negativity in the literature on ``ultra-discrete limits'' might suggest that it is the fact that $\gamma<0$ that is causing a problem here, but that would be a drastic oversimplification.    The fact is that for \textit{any} choice of $\gamma\not=-4$ the prediction of \eqref{temp} would have been correct and the limit would have the value $a$ as expected.  So, an explanation of this phenomenon would have to shed light on the question of why this occurs only at $\gamma=-4$.    (Such an explanation is provided in the following section.)

This example is not as artificial as it may at first seem since it will arise in a more general form as part of an investigation of a simple ordinary difference equation in Section~\ref{sec:ordinary}. It demonstrates that the coefficients in the linear combination of exponentials can have unexpected effects on the ``ultra-discrete limit''.  It also demonstrates that the natural generalization of \eqref{orig-max} provided by \eqref{temp} is correct \textit{generically} in the sense that it is only for rather specific choices of coefficient that it fails.

\section{A More General Limit}\label{proofsec}

This section states and proves a generalization of \eqref{orig-max} which will later be applied to questions involving difference equations.
We wish to consider instead the case in which the argument of the logarithm, $\tau$, may be an arbitrary linear combination of exponentials and in which the numerator of the exponents can also depend upon $\eps$.  The following theorem gives conditions that are sufficient (but not necessary, as we will see in Section~\ref{sec:ordexamp1}) to conclude that the limit of $\eps \ln\tau$ is still equal to the largest limit of the numerators of the exponents.

\begin{theorem}\label{main}
 Let
\begin{equation}
\tau=\sum_{i=1}^n \gamma_i e^{\frac{\wei}{\eps}}
\label{tau}
\end{equation}
where $\gamma_i\in\C$ are arbitrary, non-zero constants and $\wei$ are real valued functions
such that the limits 
$$
M_i=\lim_{\eps\to0^+}\wei \qquad i=1,\ldots,n
$$
are finite.
Let
$$
\M=\Max(M_1,\ldots,M_n) \qquad
\hbox{and}
\qquad
\II=\{i|M_i=\M\}
$$
denote the largest value of these limits and the set of indices for which this maximum is achieved, respectively.  

Then
$$
\lim_{\eps\to0^+} \eps\ln\tau=\M
$$
if either
\begin{enumerate}
\item[i.] $|\II|=1$ (the maximum is achieved only once)
\item[or] \phantom{a}
\item[ii.] the limits 
$$
\mu_i=\lim_{\eps\to0^+}(\wei-\M)/\eps \qquad\hbox{for}\qquad i\in\II
$$
are finite and $D\not=0$ where
$$
D=\sum_{i\in\II}\gamma_ie^{\mu_i}. $$

\end{enumerate}
\end{theorem}
\begin{proof}

In the case $|\II|=1$ let $\II=\{i_0\}$ so that $M_{i_0}=\M>M_i$ for $i\not=i_0$.
Then
\begin{eqnarray*}
\lim_{\eps\to0^+}\eps \ln \tau&=&
\lim_{\eps\to0^+}\eps \ln\left[ e^{\M/\eps}\sum \gamma_i e^{(\wei-\M)/\eps}\right]\\
&=& \eps \ln e^{\M/\eps} +\eps\ln \sum \gamma_ie^{(\wei-\M)/\eps}\\
&&\hbox{\tiny (then since $\wei-\M<0$ for $i\not=i_0$ and $\eps$ sufficiently small)}\\ 
&=& \M+\lim_{\eps\to0^+}\eps\ln\gamma_{i_0}=\M.
\end{eqnarray*}

The situation is more complicated in the case that $|\II|>1$.  We now assume that $D\not=0$ and show that again the limit is simply equal to the maximum of the exponents.

First, we wish to show that $D\not=0$ implies that  $\tau$ is non-zero on a small open interval $(0,\beta)$ and so it makes sense to consider the function $\eps\ln\tau$.  (There is no need to require $\tau>0$, however, since the proof to follow continues to be valid in the case that this function takes on complex values or even if it becomes multivalued as a consequence of analytic continuation.)
\begin{eqnarray*}
D&=&\sum_{i\in\II}\gamma_ie^{\mu_i}
= \lim_{\eps\to0^+}\sum_{i\in\II}\gamma_i e^{(\wei-\M)/\eps}\\
&&\hbox{\tiny (then since $\wei-\M<0$ for $i\not\in\II$ and $\eps$ sufficiently small)}\\
&=& \lim_{\eps\to0^+}\sum_{i=1}^n \gamma_ie^{(\wei-\M)/\eps}
= \lim_{\eps\to0^+}e^{-\M/\eps}\tau
\end{eqnarray*}
Suppose $\eps_i>0$ is a sequence that converges to $0$ and with the property that $\tau=0$ for $\eps=\eps_i$.  Then, since the limit above converges to $D$, it must be the case that $D=0$. 
Since $D\not=0$, no such sequence of $\{\eps_i\}$ can exist.

Now, we simply compute that
\begin{eqnarray*}
\lim_{\eps\to0^+}\eps \ln \tau&=&
\lim_{\eps\to0^+}\eps \ln\left[ e^{\M/\eps}\sum \gamma_i e^{(\wei-\M)/\eps}\right]\\
&=&\lim_{\eps\to0^+} \eps \ln e^{\M/\eps} +\lim_{\eps\to0^+}\eps\ln \sum_{i=1}^n e^{(\wei-\M)/\eps}\\
&=& \M +\lim_{\eps\to0^+}\eps\ln \left(\sum_{i\in\II} e^{(\wei-\M)/\eps}
+\sum_{i\not\in\II} e^{(\wei-\M)/\eps}\right)\\
&=& \M +\lim_{\eps\to0^+}\eps\ln \left(D
+0\right)
= \M.
\end{eqnarray*}

\end{proof}

Note that, apart from the restriction that $D\not=0$, the value of the limit is independent of the choice of coefficients $\gamma_i$.    We will take advantage of this feature in Sections~\ref{sec:ordinary}-- \ref{sec:integrals} to effectively apply the procedure of  ultra-discrete limits to integrable difference equations even in the absence of any positivity restrictions.
Furthermore, note that as a consequence of this lack of dependence on the value of the coefficients, the limit will be unaffected by an extra common factor of $-1$.  In particular, we have the following corollary:
\begin{corollary}
If the conditions of Theorem~\ref{main} are met then
$$
\lim_{\eps\to0^+}\eps\ln|\tau|=\lim_{\eps\to0^+}\eps\ln\tau.
$$
\end{corollary}
Again, this ability for us to take the absolute value prior to taking the ultra-discrete limit will be useful in our attempt to apply this result in the case of difference equations.

\subsection{Continuation of Earlier Example}

It is important to note that if neither conditions (i) or (ii) are met, then the value of the limit cannot necessarily be determined through knowledge of the limits $M_i$ of the functions in the exponents.  This is demonstrated by the example from Section~\ref{sec:examp} in which the value of the limit of the entire expression, $a-k$, depends on the parameter $k$ not appearing in the limits of the functions in the exponents when $\gamma=-4$.

We can now apply Theorem~\ref{main} to explain the significance of $\gamma=-4$ in that example.   Again let
$$
\tau=e^{\ue_1/\eps}+\gamma e^{\ue_2/\eps}
\qquad
\ue_1=a+2\eps\ln\left(e^{-k/\eps}+\sqrt{8+e^{-2k/\eps}}\right)
\qquad\hbox{and}\qquad
\ue_2=a+\eps\ln2
$$
but consider $\gamma$ arbitrary.
The paramters $\mu_i$ measuring convergence are then
$$
\mu_1=\lim_{\eps\to0^+}\left(\ue_1-a\right)/\eps=\ln 8
\qquad
\hbox{and}
\qquad
\mu_2=\lim_{\eps\to0^+}\left(\ue_2-a\right)/\eps=\ln2.
$$
Hence, we find that
$$
D=e^{\ln 8}+\gamma e^{\ln 2}=2(4+\gamma).
$$
Since this is only equal to zero at $\gamma=-4$, the ``mystery'' is resolved. 

\subsection{An Example where $D$ Does Not Exist}

To illustrate the role of condition (i) in Theorem~\ref{main}, we briefly point out that for
$$
\tau=e^{\ue_1/\eps}-e^{\ue_2/\eps}
\qquad
\ue_1=\eps\sin(1/\eps)\qquad \ue_2=-2+3\eps
$$
one has $\M=\lim_{\eps\to0^+}\ue_1=0>\lim_{\eps\to0^+}\ue_2=-2$.  Since $|\II|=1$ we know that $\lim_{\eps\to0^+}\eps\ln\tau=0$ even though the limit defining $\mu_1$ has no value.

\section{Application: Ordinary Difference Equations}\label{sec:ordinary}

%

Consider the difference equation
\begin{equation}
\sum_{i=1}^N \left(\alphae_i  \prod_{j=-d}^d (\xe_{n+j})^{m_{ij}}\right)
=
\sum_{i=1}^{\hat N} \left(\hatalphae_i \prod_{j=-d}^d (\xe_{n+j})^{\hat m_{ij}}\right).
\label{diffeq}
\end{equation}
We will generally be interested in the case $m_{ij},\hat m_{ij}\in\Z$ for the sake of preserving the integrality of solutions under the ultra-discrete limit.

\begin{definition}\label{candidate}
We say that a solution $\xe_n$  to \eqref{diffeq} is a candidate for the ultra-discrete limit if:
\begin{itemize}
\item  $\xe_{n-d},\ldots,\xe_{n+d}$,  $\alphae_i$, and $\hatalphae_i$ take \textit{non-zero} real values for $\eps\in(0,\beta)$ for some small, positive number $\beta$;

\item The limits $\displaystyle u_n=\lim_{\eps\to0^+}\eps\ln|\xe_n|$, $\displaystyle A_i=\lim_{\eps\to0^+}\eps \ln |\alphae_i|$ and
$\displaystyle \hat A_i=\lim_{\eps\to0^+}\eps \ln |\hatalphae_i|$ exist and are finite;
\item The limits
$$
\mu_i=\lim_{\eps\to0^+}\frac{(\eps\ln|\alphae_i|+\eps\sum_{j=-d}^d m_{ij}\ln| \xe_{n+j}|) -\M}{\eps} \hbox{ for }i\in\II
$$
and
$$
\hat\mu_i=\lim_{\eps\to0^+}\frac{(\eps\ln|\hatalphae_i|+\eps\sum_{j=-d}^d \hat m_{ij}\ln| \xe_{n+j}|) -\hat\M}{\eps}\hbox{ for }i\in\hat\II
$$
exist and are finite where
 $$\M=\Max_{i=1}^N(A_i+\sum_{j=-d}^d m_{ij} u_{n+j})$$ and $$\II=\{i|A_i+\sum_{j=-d}^d m_{ij} u_{n+j}=\M\}$$ is the set of indices for which this maximum is attained, and similarly,  $$\hat\M=\Max_{i=1}^{\hat N}(\hat A_i+\sum_{j=-d}^d \hat m_{ij} u_{n+j})\hbox{ and }\hat\II=\{i|\hat A_i+\sum_{j=-d}^d \hat m_{ij} u_{n+j}=\hat\M\}.$$ 
 \end{itemize}
 \end{definition}
 
\begin{theorem}\label{ordinarythm}
Suppose $\xe_n$ is a solution to the algebraic difference equation
\eqref{diffeq} which is a candidate for the ultra-discrete limit (see Definition~\ref{candidate} for terminology and notation).
The ultra-discrete limit of $\xe_n$
$$u_n=\lim_{\eps\to0^+}\eps \ln| \xe_n|$$
is a solution to the equation
\begin{equation}
\Max_{i=1}^N(A_i+\sum_{j=-d}^d m_{ij} u_{n+j})
=
\Max_{i=1}^{\hat N}(\hat A_i+\sum_{j=-d}^d \hat m_{ij} u_{n+j})
\label{udeqn}
\end{equation}
if the maximum value on each side is never attained simultaneously for more than one value of the index, or as long as both
\begin{equation}
\sum_{i\in\II}\sigma_ie^{\mu_i}\not=0 \qquad\hbox{and}\qquad \sum_{i\in\hat\II}\hat\sigma_ie^{\hat\mu_i}\not=0
\label{condition}
\end{equation}
where
$$
\sigma_i=\frac{\left|\alphae_i\prod_{j=-d}^d (\xe_{n+j})^{m_{ij}}\right|}{\alphae_i\prod_{j=-d}^d (\xe_{n+j})^{m_{ij}}}
\qquad
\hat\sigma_i=\frac{\left|\hat\alphae_i\prod_{j=-d}^d (\xe_{n+j})^{\hat m_{ij}}\right|}{\hat\alphae_i\prod_{j=-d}^d (\xe_{n+j})^{\hat m_{ij}}}
$$
are the signs of the corresponding term in the equation.
%

\end{theorem}

\begin{proof}
Letting
$$
\wei=\eps \ln\left|\alphae_i \prod_{j=-d}^d (\xe_{n+j})^{m_{ij}}\right|
$$
we can write the leftside of \eqref{diffeq} in the form
$$
\tau=\sum_{i=1}^N  \sigma_i e^{\wei/\eps}.
$$
If $|\II|=1$ or $D\not=0$, then Theorem~\ref{main} tells us that the limit of $\eps\ln\tau$ will be
$$
\Max_{i=1}^N(\lim_{\eps\to0^+} \wei).
$$
But that $D$ is non-zero is exactly the first of the two requirements in \eqref{condition} and so
\begin{eqnarray*}
\lim_{\eps\to0^+}\wei&=&\lim_{\eps\to0^+}\left( \eps\ln \alphae_i+ \eps \sum_{j=-d^d} m_{ij}\ln\xe_{n+j}\right)\\
&=& A_i+\sum_{j=-d}^dm_{ij}u_{n+j}
\end{eqnarray*}
and that $|\II|=1$ is equivalent requiring that the maximum on the lefthand side of \eqref{udeqn} is achieved for a unique value of the index.

Similarly, we conclude that the limit of $\eps$ times the log of the righthand side of the equation is $\Max_i(\hat A_i+\sum_{j=-d}^d \hat m_{ij}u_{n+j})$.  If $\xe_n$ is a solution then these two sides are always equal and hence their limits are equal as well.
\end{proof}

\subsection{Basic Example}\label{sec:ordexamp1}

Consider the difference equation
\begin{equation}
\xe_{n+1}+e^{2b/\eps}\xe_{n-1}-e^{2c/\eps}\xe_{n-1}=2e^{b/\eps}\xe_n.
\label{example1}
\end{equation}
  Let $N=3$ because there are three terms on the left, $\hat N=1$ because there is one monomial on the right,  and $d=1$ because we only see $n$ shifted up and down by this much.  The monomials on the left are given by letting 
$$\alphae_1=\sigma_1=1,\ m_{1,-1}=m_{1,0}=0.\ m_{1,1}=1,
$$
$$
\alphae_2=e^{2b/\eps},\ \sigma_2=1,\ m_{2,-1}=1,\ m_{2,0}=m_{2,1}=0
$$
and
$$
\alphae_3=e^{2c/\eps},\ \sigma_3=-1,\ m_{3,-1}=1,\ m_{3,0}=m_{3,1}=0
$$
The monomial on the right is given by letting 
$$
\hat\sigma_1=1,\ \hatalphae_1=2e^{b/\eps},\ \hat m_{1,-1}=\hat m_{1,1}=0\hbox{ and }\hat m_{1,0}=1.
$$
A solution to this equation is given by $$\xe_n=\left(e^{b/\eps}+e^{c/\eps}\right)^n$$ since expanding the lefthand side of \eqref{example1} with this definition for $\xe_n$ gives an expression equal to the righthand  side for any $n$ or $\eps$.

Trivially, this same solution $\xe_n$ \textit{also} solves the difference equation
\begin{equation}
e^{2b/\eps}\xe_{n-1}-e^{2c/\eps}\xe_{n-1}=2e^{b/\eps}\xe_n-\xe_{n+1}.
\label{example2}
\end{equation}
In fact, in most circumstances one would want to consider this equation to be completely equivalent to \eqref{example1} since one can simply add $\xe_{n+1}$ to both sides of one to make it into the other.  \textit{However}, for the purposes of this paper they are quite different.

Regardless of whether we are working with the difference equation in the form \eqref{example1} or \eqref{example2}, the solution $\xe_n=\left(e^{b/\eps}+e^{c/\eps}\right)^n$ is the same.  Then our solution $u_n$ to the ultra-discrete equation should take the form
$$
u_n=\lim_{\eps\to0^+}\eps\ln(e^{b/\eps}+e^{c/\eps})^n=n\lim_{\eps\to0^+}\eps\ln(e^{b/\eps}+e^{c/\eps})=n\Max(b,c).
$$

Ignoring condition~\eqref{condition} of Theorem~\ref{ordinarythm} we get from \eqref{example1} that $u_n$ should satisfy
\begin{equation}
\Max(u_{n+1},2b+u_{n-1},2c+u_{n-1})=b+u_n
\label{uexamp1}
\end{equation}
and from \eqref{example2} we get that it should satisfy
\begin{equation}
\Max(2b+u_{n-1},2c+u_{n-1})=\Max(b+u_n,u_{n+1}).
\label{uexamp2}
\end{equation}

However, it is \textit{not} the case that $u_n$ satisfies each of these equations regardless of the choice of $b$ and $c$.  In particular, condition \eqref{condition} may not be met and, as we will see, this causes $u_n$ not to solve \eqref{uexamp1} when $b<c$.

Look at the arguments of $\Max$ on the left hand side of \eqref{uexamp1} and consider the question of which of them are \textit{equal} to the maximum value.   If $b>c$ then the terms have the values $(n+1)b$, $(n+1)b$ and $(n-1)b+2c$ respectively so that the first two terms are equal to the maximum value.  (Using the notation of the theorem, $\M=(n+1)b$ and $\II={1,2}$.)  All of the $\mu_i$ are zero (which means that these solutions converge quickly to their limits) and so all of the exponential terms are just equal to one.   Now, we must look at the sum of $\sigma_1$ and $\sigma_2$ to make certain that it is not equal to zero.  Since $\sigma_1+\sigma_2=2\not=0$ the condition is met.  And, in fact, one can easily verify that if $b>c$ then $u_n=nb$ \textit{is} indeed a solution to \eqref{uexamp1}.  However, when $b<c$ then things are quite different.  In that case, $\M=(n+1)c$ and $\II=\{1,3\}$, but $\sigma_1+\sigma_3=0$ and so Theorem~\ref{ordinarythm} does \textit{not} predict that $u_n$ will solve the equation.  In fact, it does not because then the lefthand side has the value $(n+1)c$ while the rightside has the value $nc+b$.

In contrast, the theorem predicts that $u_n$ will solve \eqref{uexamp2} when $b>c$ \textit{and} when $b<c$.  In particular, if $b>c$ then $\M=\Max(2b+u_{n-1},2c+u_{n-1})=(n+1)b$ and $\II=\{1\}$ while if $c>b$ then $\M=(n+1)c$ and $\II=\{2\}$.  Thus, we conclude from the theorem that $u_n$ should satisfy \eqref{uexamp2} as long as $b\not=c$.  In fact, it can be checked explicitly that $u_n$ is a solution to \eqref{uexamp2} in the case b=c as well. This is true even though the conditions
of the theorem fail to be satisfied in that case. This shows that the conditions of the theorem, while sufficient, are not necessary.

\subsection{Another example: Demonstrating the Role of Rate of Convergence}

Consider the difference equation
$$
\xe_{n+1}-2\xe_{n-1}=e^{\lambda/\eps}\xe_n.
$$
Considering the survival of solutions of this equation under the ultra-discrete limit illustrates the importance of the parameters $\mu_i$ which measure the rate of convergence of the terms.

This has a solution of the form 
$$
\xe_n=\left(\frac{\sqrt{8+e^{\frac{2 \lambda }{\eps }}}+e^{\frac{\lambda
   }{\eps }}}{2}\right)^n.
$$
Consequently, we would expect
$$
u_n=\lim_{\eps\to0^+}\eps\ln\xe_n=\left\{\begin{matrix}n\lambda&\lambda>0\\
0&\lambda\leq0\end{matrix}\right.
$$
to solve
$$
\Max(u_{n+1},u_{n-1})=\lambda+u_n.
$$
And indeed, if $\lambda>0$ this is true.  However, it fails when $\lambda<0$.  This could be predicted by the Theorem.  Condition \eqref{condition} is trivially met when $\lambda>0$ because the exponents on the left do not have the same limit.  However, when $\lambda<0$ then both exponents go to $0$ and $\mu_1=\ln 2$ and hence the sum is $D=e^{\ln 2}-e^{\ln 2}=0$.

\section{Application: Hirota Bilinear Difference Equation}\label{sec:hbde}


The Hirota Bilinear Difference Equation (HBDE) \cite{Miwa,Sato,Zabrodin} for a function $\xe(a,b,c,d)$ of four variables is
\begin{eqnarray*}
&&(\lame_4-\lame_3)(\lame_2-\lame_1)\xe(a,b,c+1,d+1)\xe(a+1,b+1,c,d)
\\&+&(\lame_4-\lame_1)(\lame_3-\lame_2)\xe(a+1,b,c,d+1)\xe(a,b+1,c+1,d)
\\&=&
(\lame_4-\lame_2)(\lame_3-\lame_1)\xe(a+1,b,c+1,d)\xe(a,b+1,c,d+1).
\end{eqnarray*}
As a consequence of the main result in \cite{GK}, a solution to this equation can be constructed as
$$
\xe(a,b,c,d)=\det\left(P\cdot (I-\lame_1S)^a\cdot (I-\lame_2S)^b\cdot(I-\lame_3S)^c\cdot(I-\lame_4S)^d \cdot \Omega\right)
$$
where $\Omega$ is any $n\times k$ matrix ($k<n$), $S$ is any $n\times n$ matrix such that the lower-left $k\times(n-k)$ block has rank one, $I$ is the $n\times n$ identity matrix, and $P=(0\ I_k)$ is the $k\times n$ matrix which has the $k\times k$ identity matrix as a block and all zeroes to the left of it.

It is interesting to note that even though $\xe(a,b,c,d)$ has dependence upon $\eps$ inheritted from the presence of $\lame_i$ in the formula, one is also free to choose $S$ and $\Omega$ to depend upon $\eps$ and $\xe(a,b,c,d)$ would still solve the difference equation.

For example, let us consider
$$
\lame_1=e^{-2/\eps}-e^{2/\eps}+e^{3/\eps}
\qquad
\lame_2=e^{3/\eps}-e^{2/\eps}
\qquad
\lame_3=e^{3/\eps}+e^{-2/\eps}
\qquad
\lame_4=e^{3/\eps}
$$
so that 
$$
(\lame_4-\lame_3)(\lame_2-\lame_1)=e^{-4/\eps}
\qquad
(\lame_4-\lame_2)(\lame_3-\lame_1)=e^{4/\eps}.
$$
(It is necessarily the case that $(\lame_4-\lame_1)(\lame_3-\lame_2)$ is the difference of the other two.)

Now, just for the sake of example, let us consider the solution which comes from using $k=2$, $n=4$, 
$$
S=
\left(
\begin{array}{llll}
 1 & 0 & 0 & 0 \\
 1 & e^{\relax{1/\eps}} & 0 & 0 \\
 0 & 0 & e^{5/\eps} & 0 \\
 0 & e^{5/\eps} & 0 & e^{12/\eps}
\end{array}
\right)
\qquad\hbox{and}\qquad
\Omega=
\left(
\begin{array}{ll}
 0 & 0 \\
 1 & 0 \\
 1 & 1 \\
 0 & 1
\end{array}
\right).
$$
This solution has the form\footnote{Up to a common multiple independent of $a$, $b$, $c$ and $d$ which does not matter due to the bilinear nature of the equation.}
$$
\xe(a,b,c,d)=\xe_1(a,b,c,d)-\xe_2(a,b,c,d)+\xe_3(a,b,c,d)
$$
where
\begin{eqnarray*}
\xe_1(a,b,c,d)&=&
e^{4/\eps } \left(1-e^{4/\eps }\right)^d
   \left(1-e^{-1/\eps }-e^{4/\eps }\right)^c
   \left(1+e^{3/\eps }-e^{4/\eps }\right)^b
   \left(1-e^{-1/\eps }+e^{3/\eps }-e^{4/\eps
   }\right)^a 
   \\&&
   \left(1-e^{8/\eps }\right)^d
   \left(1-e^{3/\eps }-e^{8/\eps }\right)^c
   \left(1+e^{7/\eps }-e^{8/\eps }\right)^b
   \left(1-e^{3/\eps }+e^{7/\eps }-e^{8/\eps
   }\right)^a
   \end{eqnarray*}
   \begin{eqnarray*}
   \xe_2(a,b,c,d)&=&
   \left(1-e^{8/\eps }\right)^d \left(1-e^{3/\eps
   }-e^{8/\eps }\right)^c \left(1+e^{7/\eps
   }-e^{8/\eps }\right)^b \left(1-e^{3/\eps
   }+e^{7/\eps }-e^{8/\eps }\right)^a
   \\&&
   \left(1-e^{15/\eps }\right)^d
   \left(1-e^{10/\eps }-e^{15/\eps }\right)^c
   \left(1+e^{14/\eps }-e^{15/\eps }\right)^b
   \left(1-e^{10/\eps }+e^{14/\eps
   }-e^{15/\eps }\right)^a
   \end{eqnarray*}
   and
   \begin{eqnarray*}
   \xe_3(a,b,c,d)&=&
   e^{11/\eps } \left(1-e^{8/\eps }\right)^d
   \left(1-e^{3/\eps }-e^{8/\eps }\right)^c
   \left(1+e^{7/\eps }-e^{8/\eps }\right)^b
   \left(1-e^{3/\eps }+e^{7/\eps }-e^{8/\eps
   }\right)^a 
     \\&&
     \left(1-e^{15/\eps }\right)^d
   \left(1-e^{10/\eps }-e^{15/\eps }\right)^c
   \left(1+e^{14/\eps }-e^{15/\eps }\right)^b
   \left(1-e^{10/\eps }+e^{14/\eps
   }-e^{15/\eps }\right)^a.
   \end{eqnarray*}
   
Before even worrying about the structure of the equation, one must worry that this solution itself will
   not have a well defined or simple ultra-discrete limit.  In particular, if we define
   $$
   u(a,b,c,d)=\lim_{\eps\to0^+}\eps \ln \xe(a,b,c,d),
   $$
   and we do not worry about the requirement that $D\not=0$ in Theorem~\ref{main}, then we would conclude that
  \begin{equation}
  \label{uabcd}
   u(a,b,c,d)=\Max(4+12a+12b+12c+12d,
   23a+23b+23c+23d,
   11+23a+23b+23c+23d).
   \end{equation}
   (This is found by adding up for each factor in the expression of $\xe_i$ with $i=1,2,3$ the highest numerator of the exponent when viewed as a fraction having $\eps$ as the denominator.  For instance, the first expression is found from $\xe_1$ as $4+4d+4c+4+4a+8d+8c+8b+8a$.)
      Yet, as the examples of the previous section demonstrate, we ought to be worried that this formula is not accurate in the case that $D=0$ when Theorem~\ref{main} is applied to this example.

In fact, it turns out that there is no need to worry about this since it is always the case here that $|\II|=1$.
Note first that the third is always $11$ more than the second, and so there is no possibility that these would be equal.  By the same argument, there would be no problem if the first two were equal since they could not be the maximum (with the third being $11$ more all of the time).  Finally, we need only worry about the possibility that the first and third are equal.  Certainly, it \textit{is} possible for these two expressions to have the same value, but \textit{never for integer values of $a,b,c,$ and $d$.}  In particular, they are only equal when 
$$
a=-\left(b+c+d+\frac{7}{11}\right).
$$

A similar argument applies when we wish to compute the limit of the Hirota Bilinear Difference Equation to find the equation that this $u(a,b,c,d)$ satisfies.   Rewriting the HBDE as
\begin{eqnarray*}
\sigma_1 e^{(\ue(a+1,b+1,c,d)+\ue(a,b,c+1,d+1)-4)/\eps}
&+&
\sigma_2 e^{(\ue(a+1,b,c,d+1)+\ue(a,b+1,c+1,d)+\eps\ln(e^{4/\eps}-e^{-4/\eps}))/\eps}\\
&=&\sigma_3 e^{(\ue(a+1,b,c+1,d)+\ue(a,b+1,c,d+1)+4)/\eps}
\end{eqnarray*}
where $\ue(a,b,c,d)=\eps\ln |u(a,b,c,d)|$ and $\sigma_1=\pm1$ as needed when $u$ is negative, leads to its ultra-discretization as 
$$
\Max\left(u(a+1,b+1,c,d)+u(a,b,c+1,d+1)-4,\
u(a+1,b,c,d+1)+u(a,b+1,c+1,d)+4\right)$$
\begin{equation}
\label{UDHBDE}
=
u(a+1,b,c+1,d)+u(a,b+1,c,d+1)+4.
\end{equation}
Then to show that \eqref{uabcd} gives a solution we apply Theorem~\ref{ordinarythm} and note that since
$$
u(a+1,b+1,c,d)=u(a,b,c+1,d+1)=u(a+1,b,c,d+1)=u(a,b+1,c+1,d)
$$
the first argument of the $\Max$ operator is always $8$ less than the second argument.


\section{Application: First Integrals}\label{sec:integrals}

In this section, we consider difference equations that admit first integrals and ask the following question:
When does the ultra-discretization of the equation leave the ultra-discretization of the first integral invariant? The study performed in this section   will be more algebraic than analytic: the results about
first integrals will apply to the complete set of solutions as opposed to a particular solution as considered before.

\subsection{Preliminary examples}
\label{sec:preint}

A well-known class of second order difference equations is given by the {\sl{QRT}} mappings discovered by Quispel, Roberts, and Thompson  \cite{QuRoTh89}. The ultra-discrete limit
of members of that class of mappings together with their first integrals was first performed in \cite{TaToGrOhRa97} but, each time this is done, one has to check explicitly  that the resulting ultra-discretization of the first integral is indeed invariant. 

\newcommand{\xuu}{{x}_{n+2}}
\newcommand{\xu}{{x}_{n+1}}
\newcommand{\xd}{{x}_{{n-1}}}
\newcommand{\x}{{x}_{n}}
\newcommand{\Xuu}{{{u}_{n+2}}}
\newcommand{\Xuuu}{{{{u_{n+3}}}}}
\newcommand{\Xuuuu}{{{{u_{n+4}}}}}
\newcommand{\Xuuuuu}{{{{u_{n+5}}}}}
\newcommand{\Xuuuuuu}{{{{u_{n+6}}}}}
\newcommand{\Xu}{{u}_{n+1}}
\newcommand{\Xd}{{u}_{{n-1}}}
\newcommand{\X}{{u}_{n}}
\newcommand{\m}{{\Max}}
\def\iotae{\iota^{\eps}}
\def\betae{{\beta}^{\eps}}

Let us start with the following simple example of a  difference equation 
\begin{equation}
\xu=\frac{\alphae\x+1}{\x\xd},
\label{dPI-2}
\end{equation}
 where $\alphae$ is constant in $n$. Equation (\ref{dPI-2}) is a member of the class of QRT 
 mappings. As such, (\ref{dPI-2}) admits the following first integral
\begin{equation}
\iotae(\xd,\x)=\x+\xd+\frac{\alphae}{\x}+\frac{\alphae}{\xd}+\frac{1}{\x\,\xd}.
\label{I}
\end{equation}
We say that $\iotae$ is a first integral for \eqref{dPI-2} because $\iotae(\xd,\x)=\iotae(\x,\xu)$ if $\x$ is a solution to \eqref{dPI-2}.
To obtain the ultra-discrete version of these two expressions, one writes $\alphae=e^{A/\eps}$ and $\x=e^{\X/\eps}$.
Then we take the limit as $\epsilon\to0^+$ of the equation using the identity (\ref{orig-max}) (note that in this particular example, 
we do not need the more general relation given by Theorem \ref{main}).
We obtain
 \begin{equation}
\Xu=\m(\X+A,\;0)-\X-\Xd
\label{udPI2a}
\end{equation}
and
\begin{equation}
\label{IUD}
I=\m\bigl(\X,\;\Xd,\;A-\Xd,\;A-\X,\;-\X-\Xd\bigr).
\end{equation}
The quantity $I$ defined in (\ref{IUD}) is
conserved by the equation (\ref{udPI2a}). Following \cite{JoLa06}, this can be shown directly by substituting 
the expression for $\Xd$ in terms of $\X$ and $\Xu$ coming from (\ref{udPI2a})
into the definition  of $I$ as follows:
\begin{eqnarray*}
I_n
&=&\m\Bigl(\X,\;
-\Xu-\X+(\X+A)_+,\;
A-\X,\\
&&\qquad\qquad
A+\Xu+\X-(\X+A)_+,\;
\Xu-(\X+A)_+\Bigr)\\
&=&\m\Bigl(\X,\,A-\Xu,\;-\X-\Xu,\;A-\X,\\
&&\qquad\quad
\m\bigl(A+\Xu+\X-(\X+A)_+,\;\Xu-(\X+A)_+\bigr)\Bigr)\\
&=&\m\left(\X,\,A-\Xu,\;-\X-\Xu,\;A-\X,\;
,\;\Xu\right)\\
&=&
I_{n+1},
\end{eqnarray*}
where we have introduced the notation $(k)_+\equiv \m\left(k,\;0\right)$. 

Another more complex example is given by a difference equation that is related (through the process of deautonomisation) to a discrete version of the third Painlev\'e equation 
\cite{RaGr96,RaTaGrOh98,GrNiRa99} 
\begin{equation}
\xu=\frac{\left(\alphae_1\alphae_2+\x\right)\left({\alphae_1}/{\alphae_2}+\x\right)}
{\xd\left(\alphae_1\alphae_3\x+1\right)\left({\alphae_1}/{\alphae_3}\x+1\right)}.
\label{P6a}
\end{equation}
Again, as a member of the QRT family of mapping, \eqref{P6a} has a first integral given by
\begin{equation}
\begin{aligned}
\iotae&=
\alphae_{{1}}{\alphae_{{3}}}^{}(1+\alpha^{{\eps 2}}_2)\left( \frac{1}{x_{n-1}}+ \frac{1}{x_{n}} \right)
 + \alphae_{{1}} \alphae_{{2}}(1+ \alpha_{{3}}^{\eps 2}) \left( x_{{n}}+x_{{n-1}}
 \right)\\&
  + 
{\alphae_{{1}}}^{2} \alphae_{{2}} \alphae_{{3}}   \left( x_{{n}}x_{{n-1}}+{\frac {1}{x_{{n}}x_{{n-1}}}} \right) +
 \alphae_{{2}} \alphae_{{3}} \left({\frac {x_{{n}}}{x_{{n-1}}}}+{\frac {x_{{n-1}}}{x_{{n}}}} \right).
\end{aligned}
\end{equation}
The ultra-discretization of these two expressions is performed the same way as before by setting 
$\alphae_i=e^{A_i/\epsilon}$ and $\x=e^{\X/\eps}$. In the limit as $\eps\rightarrow 0^+$, one gets
\begin{eqnarray}
\label{udP6a}
\Xu&=&2\X+(A_1+A_2-\X)_+\\
\nonumber&&+(A_1-A_2-\X)_+-(A_1+A_3+\X)_+\\
\nonumber&&-(A_1-A_3+\X)_+-\Xd,
\end{eqnarray}
with first integral
\begin{eqnarray}
\label{IUPII2a}I&=&\m\Bigl(A_1+A_3-\Xd,\;A_1+A_3-\X,\;\\
\nonumber&&\qquad \;A_1+A_3+2A_2-\Xd,\;A_1+A_3+2A_2-\X,\;\\
\nonumber &&\qquad 
A_1+A_2+\X,\;A_1+A_2+\Xd,
\\
\nonumber&&\qquad 
A_1+A_2+2A_3+\X,\;A_1+A_2+2A_3+\Xd,\\
\nonumber&&\qquad 2A_1+A_2+A_3+\X+\Xd,\;2A_1+A_2+A_3-\X-\Xd,\;\\
\nonumber&&\qquad A_2+A_3+\X-\Xd,\;A_2+A_3+\Xd-\X\;\Bigr).\nonumber
\end{eqnarray}
Checking that $I$ defined above is a first integral is quite a cumbersome 
task. Instead, we will rely on a general theorem
that we prove below.

\subsection{Main result about first integrals}

Consider the difference equation
\begin{equation}
\xe_{n+d}
=
f(\xe_{n},\xe_{n+1},...,\xe_{n+d-1})
\label{evdiffeq}
\end{equation}
where $f$ is a rational function on $\mathbb{R}^d$ of the form
$$
f(a_0,a_1,...,a_{d-1})\equiv 
\frac{\sum_{i=1}^N \left(\alphae_i  \prod_{j=0}^{d-1} (a_{j})^{m_{ij}}\right)}{\sum_{i=1}^{\hat N} \left(\hatalphae_i \prod_{j=0}^{d-1} (a_{j})^{\hat m_{ij}}\right)}.
$$

A first integral for (\ref{evdiffeq})  is defined to be a rational function $\iotae$ on $\mathbb{R}^d$ 
of the form 
\begin{equation}
\iotae(a_0,a_1,...,a_{d-1})=\frac{\sum_{i=1}^M \left(\betae_i  \prod_{j=0}^{d-1} (a_{j})^{l_{ij}}\right)}{
\sum_{i=1}^{\hat{M}} \left(\hat{\beta}^\eps_i  \prod_{j=0}^{d-1} (a_{j})^{\hat{l}_{ij}}\right)},
\label{const}
\end{equation}
where $l_{ij},\hat{l}_{ij}\in \N$, which remains invariant when restricted to any solution to (\ref{evdiffeq}), that is, $\iotae(a_1,a_2,...,a_{d-1},f(a_0,a_1,...,a_{d-1}))=\iotae(a_0,a_1,...,a_{d-1})$. Here, we assume the coefficients $\betae_i$ and $\hat{\beta}^\eps_i$ to be nonzero for $\eps$ in a certain interval $(0,\beta)$. In all our examples, these coefficients will be exponentials and thus never zero.

The following question arises: is the ultra-discretization of  (\ref{const}) a first integral for the ultra-discretization of (\ref{evdiffeq})? Precisely, does the function
$$
I(b_0,b_1,...,b_{d-1})=
\Max_{i=1}^M(B_i+\sum_{j=0}^{d-1} l_{ij} b_{j})
-\Max_{i=1}^{\hat{M}}(\hat{B}_i+\sum_{j=0}^{d-1} \hat{l}_{ij} b_{j}),
$$
where $B_i,\hat{B}_i$ are defined as $\displaystyle{B_i=\lim_{\eps\to0^+}\eps \ln \left|\betae_i\right|}$ and 
$\displaystyle{\hat{B}_i=\lim_{\eps\to0^+}\eps \ln \left|\hat{\beta}^\eps_i\right|}$,
define a first integral for the equation
\begin{equation}
u_{n+d}=g(u_{n},u_{n+1},...,u_{n+d-1})
\label{evudeqn}
\end{equation}
where $g$ a function on $\mathbb{R}^d$ defined by
$$
g(b_0,b_1,...,b_{d-1})\equiv 
\Max_{i=1}^N(A_i+\sum_{j=0}^{d-1} m_{ij} b_{j})-
\Max_{i=1}^{\hat N}(\hat A_i+\sum_{j=0}^{d-1} \hat m_{ij} b_{j}),
$$
where $A_i$ are defined in Definition (\ref{candidate}). 

Before stating the theorem, let us define the function $D$ on $\mathbb{R}^d$ by
$$
D(b_0,b_1,...,b_{d-1})\equiv \sum_{i\in\II}\sigma_ie^{\mu_i} 
$$
where
$$
\sigma_i=\frac{\left|\betae_i \left(f(e^{b_0/\eps},e^{b_1/\eps},...,e^{b_{d-1}/\eps})\right)^{l_{id-1}}\right|}{\betae_i \left(f(e^{b_0/\eps},e^{b_1/\eps},...,e^{b_{d-1}/\eps})\right)^{l_{id-1}}}
$$
($\sigma_i$ is the sign of 
$\betae_i \left(f(e^{b_0/\eps},e^{b_1/\eps},...,e^{b_{d-1}/\eps})\right)^{l_{id-1}}$
for $\eps$ close to zero)
and
$$\M=\Max_{i=1}^M(B_i+\sum_{j=0}^{d-2} l_{i{j}} b_{j+1}+l_{id-1}g(b_0,b_1,...,b_{d-1}))$$ and 
$$\II=\{i|B_i+\sum_{j=0}^{d-2} l_{i{j}} b_{j+1}+l_{id-1}g(b_0,b_1,...,b_{d-1})=\M\}.$$
and
$$
\mu_i=\lim_{\eps\rightarrow 0^+}\frac{\eps \ln \left|\betae_i\right|+\sum_{j=0}^{d-2} l_{i{j}} b_{j+1}+l_{id-1}h^{\eps}(b_0,b_1,...,b_{d-1})-\M}{\eps} \hbox{ for }i\in\II,
$$
where 

$$
h^{\eps}(b_0,b_1,...,b_{d-1})=\eps\ln{\left(\left|f(e^{b_0/\eps},e^{b_1/\eps},...,e^{b_{d-1}/\eps})\right|\right)}.
$$

Note that
$$
\lim_{\eps\rightarrow 0^+} h^{\eps}(b_0,b_1,...,b_{d-1})=g(b_0,b_1,...,b_{d-1}).
$$

Furthermore, the function $\hat{D}$ is defined the same way by replacing $\sigma_i,\mu_i,\alphae_i,B_i,l_{ij},M$ above by 
$\hat{\sigma}_i,\hat{\mu}_i,\hat{\alpha}^{\eps}_i,\hat{B}_i,\hat{l}_{ij},\hat{M}$

\begin{theorem}\label{intthe}
If $D$ and $\hat{D}$ are nonzero almost everywhere in $\mathbb{R}^d$,
 then $I$ is a first integral for 
(\ref{evudeqn}).
\end{theorem}
\begin{proof}
By definition of $I$, we have the equality
$$
\lim_{\eps \rightarrow 0^+} \eps\ln{\left(\iotae\left(e^{b_0/\eps},e^{b_1/\eps},...,e^{b_{d-1}/\eps}\right)\right)}=
I(b_0,b_1,...,b_{d-1}).
$$
Furthermore, as a direct consequence of Theorem \ref{main}, since both $D$ and $\hat{D}$ are nonzero almost everywhere in $\mathbb{R}^d$,
the following equality must hold everywhere except in a subset of measure zero
$$
\lim_{\eps \rightarrow 0^+}
\eps 
\ln{\left( \iotae\left(e^{b_1/\eps},e^{b_2/\eps},...,e^{b_{d-1}/\eps},
e^{h^{\eps}(b_0,b_1,...,b_{d-1})}\right)\right)}=
I(b_1,b_2,...,b_{d-1},g(b_0,b_1,...,b_{d-1})).
$$
Because both the argument of the limit and the right hand side of the equation are continuous in $\mathbb{R}^d$, the equality holds in all  $\mathbb{R}^d$. 

Since $\iotae$ is a first integral for \eqref{evdiffeq}, the arguments of the two limits above are equal. Hence, since both limits exist, the right sides are equal which shows that $I$ is a first integral for \eqref{evudeqn}.
\end{proof}

\subsection{Examples}

Consider the difference equation
\begin{equation}
\xuu=-\x-\xu+\frac{\alphae}{\xu}
\label{ex1}
\end{equation}
which admits the first integral given by
\begin{equation}
\iotae=\xu^2\x+\xu\x^2-\alphae(\xu+\x).
\end{equation}
(note that \eqref{ex1} was intentionally written so that $\xuu$  is a function 
of $\x$ and $\xu$ in order to apply Theorem \ref{intthe} directly with $d=2$). Equation \eqref{ex1} is related to a discrete version of the first Painlev\'e equation \cite{RaGr96,RaTaGrOh98,GrNiRa99}.
For the ultra-discrete limit we choose $\alphae=e^{A/\eps}$ and find
\begin{equation}
\Xuu=\m\left(\X,\;\Xu,\;A-\Xu\right)\equiv g(\X,\Xu),
\label{udex1}
\end{equation}
\begin{equation}
I(\X,\Xu)=\m\left(2\Xu+\X,\;\Xu+2\X,\;A+\Xu,\;A+\X\right).
\label{udintex1}
\end{equation}
The quantity defined in \eqref{udintex1} is not a first integral 
for \eqref{udex1}. To see this we consider the specific example with $A<8$, in which we have that $I(3,5)=13$ and $I(5,g(5,3))=I(5,5)=15\neq I(3,5)$.

We now show that the conditions of Theorem \ref{intthe} are not met. Specifically, it is not true that the function $D$ is nonzero almost everywhere in $\mathbb{R}^2$.
We first set
$$\betae_1=1,\ l_{1,0}=2,\ l_{1,1}=1,
$$
$$
\betae_2=1,\ l_{2,0}=1,\ l_{2,1}=2,
$$
$$
\betae_3=-e^{A/\eps},\ l_{3,0}=1,\ l_{3,1}=0,
$$
$$
\betae_4=-e^{A/\eps},\ l_{4,0}=0,\ l_{4,1}=1.
$$
and
$$
\M=\m\left(2g(b_1,b_2)+b_2,\; g(b_1,b_2)+2b_2,\;A+g(b_1,b_2),\;A+b_2\right).
$$
Furthermore, to apply Theorem \ref{intthe}, it is useful to write the following quantity explicitly
$$
f(e^{b_1/\eps},e^{b_2/\eps})=-e^{b_1/\eps}-e^{b_2/\eps}+e^{(A-b_2)/\eps}.
$$
Consider the case where $b_2$ is greater than both $b_1$ and $A-b_1$. We obtain $g(b_1,b_2)=b_2$ and $\II=\left\{1,2\right\}$. The sign of $f(e^{b_1/\eps},e^{b_2/\eps})$ for small positive $\eps$ is negative which means that $\sigma_1=-\sigma_2=1$. Since
$
\mu_1=\mu_2=0,
$
we finally get that $D=e^{\mu_1}-e^{\mu_2}=0$.  Thus the function $D$ is zero in the semi-infinite triangular region of the plane $b_2>b_1$ and $b_2>A-b_1$.
The conditions of the Theorem \ref{intthe} are thus not met.

Let us now consider the two examples of Section \ref{sec:preint}. In both cases, all the $\sigma_i$'s are equal to one. Thus, this automatically implies that both $D$ and $\hat{D}$ are nonzero. In particular, this shows that  \eqref{IUPII2a} is indeed a first integral for \eqref{udP6a}.

We now consider a last example
\begin{equation}
\xuu=\frac{\xu}{1-\xu^2}-\x\equiv f(\x,\xu)
\end{equation}
with first integral
\begin{equation}
\iota=\xu^2+\x^2-\xu^2\x^2-\xu\x.
\end{equation}
The ultra-discretization reads
\begin{equation}
\Xuu=\m\left(\X+2\Xu,\;\Xu,\;\X\right)-\m\left(2\Xu,\;0\right)\equiv g(\X,\Xu)
\label{udtricky}
\end{equation}
\begin{equation}
I=\m\left(2\Xu,\;2\X,\;2\Xu+2\X,\;\Xu+\X\right).
\label{udtrickyint}
\end{equation}
To apply Theorem \ref{intthe}, we set 
$$\betae_1=1,\ l_{1,0}=0,\ l_{1,1}=2,
$$
$$
\betae_2=1,\ l_{2,0}=2,\ l_{2,1}=0,
$$
$$
\betae_3=-1,\ l_{3,0}= l_{3,1}=2,
$$
$$
\betae_4=-1,\ l_{4,0}= l_{4,1}=1.
$$
and 
$$
\M=\m\left(2g(b_1,b_2),\;2b_2,\;2g(b_1,b_2)+2b_2,\;g(b_1,b_2)+b_2\right).
$$
Furthermore, it is useful to write the following explicitly
$$
f(e^{b_1/\eps},e^{b_2/\eps})=\frac{e^{b_2/\eps}}{1-e^{2b_2/\eps}}-e^{b_1/\eps}.
$$

To prove that the function $D$ is nonzero almost everywhere in $\mathbb{R}^2$, we consider four cases.
First, in the case $b_1,b_2>0$, we have that $g(b_1,b_2)=b_1$ and $\II=\left\{3\right\}$ and thus, because there is only one term in $D$, it cannot be zero. In the case $b_2>0,b_1<0$, $g(b_1,b_2)=\m(b_1,\;-b_2)$ and $\II=\left\{2\right\}$. Again, $D$ cannot be zero. In the case $b_2<0,b_1>b_2$, we have that $g(b_1,b_2)=b_1$ and $\II=\left\{1\right\}$. Finally, in the case $b_1<b_2<0$, $g(b_1,b_2)=b_2$ and $\II=\left\{1,2,4\right\}$. Here, $D$ is not trivially nonzero and a little more work is required. The sign of $f(e^{b_1/\eps},e^{b_2/\eps})$ for small $\eps$ is positive, $\mu_i=0,\;i=1,2,4$ and $\sigma_1=\sigma_2=-\sigma_4=1$. Thus, $D=1+1-1=1\neq 0$. We thus have proven that $D$ is nonzero almost everywhere in $\mathbb{R}^2$ which means, by Theorem \ref{intthe} that \eqref{udtrickyint} is a first integral for \eqref{udtricky}. Note however that $D$ is not nonzero everywhere in $\mathbb{R}^2$. For example, in the case $b_1=0$ and $b_2>0$, we have that $g(b_1,b_2)=0$, $\II=\left\{2,3\right\}$, the sign of  $f(e^{b_1/\eps},e^{b_2/\eps})$ is negative, $\sigma_2=-\sigma_3=1$, and thus $D=0$.

\section{Conclusions}

Although it is the case that problems can arise in computing a limit of the form 
$$
\lim_{\eps\to0^+}\eps\ln\tau
$$
with $\tau$ of the form \eqref{tau} when some terms are negative, it is not true that these problems arise ``most of the time''.  Consequently, it is often still possible to use the procedure of ultra-discrete limits developed in \cite{TTMS} to produce integer solutions to equations involving the ``Max'' operator even in the absence of positivity, as illustrated in Sections~\ref{sec:ordinary} and \ref{sec:hbde}.  Furthermore, it is especially useful to apply the results of Theorem~\ref{main} to the question of which integrals of motion are preserved by an ultra-discrete limit since in that case one does not need to consider individual solutions (cf. Section~\ref{sec:integrals}).

 It should be noted that there are various different possibilities for the value of the limit in the case that $D=0$ and $|\II|>1$ which we do not presently have the ability to differentiate in general.  For instance, in the example presented in Section~\ref{sec:ordexamp1} the value of the limit is not the maximum of the exponents as usual but rather the next largest of the exponents.  (This happens here because the sum that forms $D$ does not only vanish in the limit but actually is equal to zero for all $\eps$, as briefly mentioned in \cite{IGRS}.)  On the other hand, in the case of the limit
$$
\lim_{\eps\to0^+} \eps\ln\left(e^{\ue_1/\eps}-e^{\ue_2/\eps}+e^{\ue_3/\eps}\right)
$$
with $\ue_1=\sin(2\eps)/\eps$, $\ue_2=2\sin(\eps)/\eps$ and $\ue_3=1$, the limit yields the maximum value of $2$ despite the fact that $D=0$.  Most interestingly, in the example shown in Section~\ref{sec:examp} it is evident that the value of the limit is not determined by the values of the limits of the functions in the exponents, since the ultra-discrete limit, $a-k$, could be varied arbitrarily by selecting $k$ without affecting the limits of the individual exponents, all equal to $a$.
It would be interesting to extend Theorem~\ref{main} to be able to predict the value of the limit in the case that neither conditions (i) nor (ii) are met. 
%

%
\par\bigskip\par\noindent{\textit{Acknowledgements:}} We are grateful to our colleagues Herb Silverman and Tom Kunkle for helpful discussions and advice.
\def\mybibitem[#1]#2{\bibitem{#2}}

\end{document}